\def\der#1#2{{\partial#1\over\partial#2}}
\def\dder#1#2#3{{\partial^2#1\over{\partial#2\partial#3}}}

\documentclass[12pt,a4paper]{iopart}
\usepackage{iopams}
\usepackage{cite}
\begin{document}
\eqnobysec

\title[Equivalence transformations]{Equivalence transformations
and differential invariants of a generalized nonlinear Schr\"{o}dinger
equation}

\author{M. Senthilvelan\dag, M. Torrisi\ddag \hskip6pt and A. Valenti\ddag}

\address{\dag Centre for Nonlinear
Dynamics, Department of Physics, Bharathidasan \\
University, Tiruchirapalli 620 024, Tamil Nadu, India}

\address{\ddag Dipartimento di Matematica e Informatica,
Universit\`a di Catania  \\
Viale A. Doria 6, 95125, Catania, Italy}

\date{}

\begin{abstract}
By using the Lie's invariance infinitesimal criterion we obtain the
continuous equivalence transformations of a class of  nonlinear
Schr\"{o}dinger equations with variable coefficients. Starting from
the  equivalence generators we construct the differential invariants
of order one. We apply these latter ones to find the most general
subclass of variable coefficient nonlinear Schr\"{o}dinger equations
which can be mapped, by means of an equivalence transformation, to
the well known cubic Schr\"{o}dinger equation.  We also provide the
explicit form of the transformation.
\end{abstract}

%\submitto{\JPA}

\maketitle

\section{Introduction}

In this paper we investigate the equivalence transformations and
the differential invariants associated with the following
$(1+1)$ dimensional generalized nonlinear Schr\"{o}dinger equation
\begin{equation}
iu_t+f(t,x)u_{xx}+g(t,x)|u|^2u+h(t,x)u = 0,
\label{nlsorg}
\end{equation}
where $f(t,x),g(t,x)$ and $h(t,x)$ are real functions of $t$ and
$x$ and subscript denotes partial differentiation with respect to that
variable.  In the case $f,g=1$ and $h=0$, Eq.~(\ref{nlsorg})
reduces to the well known nonlinear Schr\"{o}dinger equation (NLS)
\begin{equation}
iu_t+u_{xx}+|u|^2u = 0.
\label{snls}
\end{equation}
Eq.~(\ref{snls}) arises
in several branches of physics including Nonlinear Optics, Condensed
Matter Physics, Hydrodynamics and so on (for the derivation and wide
applications of Eq.~(\ref{snls}), refer
\cite{Ablowitz:1,Sulem,Lak,Drazin,Ablowitz:2}).
It has been shown that the NLS equation is one of
the completely integrable nonlinear partial differential equations (PDEs)
in (1+1) dimensions and admits several interesting mathematical properties
including
infinite number of conservation laws, Lie-B\"{a}cklund symmetries,
$N$-soliton solutions and so on.
\par
An equivalence transformation for the family ~(\ref{nlsorg}) is a
non-degenerate transformation of dependent and independent
variables mapping (\ref{nlsorg}) to another equation of the same
family but with different functions, say
$(\hat{f}$,$\hat{g}$,$\hat{h})$, from the original ones.  Thus
solutions of an equation can be transformed to the solutions of an
equivalent equation.  The main advantage of this procedure is that
instead of solving individual equations one can develop schemes
for complete equivalent classes.  One may note that the classical
Lie symmetry transformations are nothing but special subgroups of
these equivalence groups of transformations since Lie symmetry transformations
map an equation into itself.  Recently, several works have been
devoted to study
the equivalence transformations of certain important nonlinear
dynamical systems (see for a review Refs.~
\cite{OV,OL,I:1} and bibliography therein).
\par
One of the classical studies in the theory of differential equations is finding
differential invariants.  The origin dates back to Laplace who derived two
invariants for linear hyperbolic equations.  These invariants are now called
Laplace invariants.  The generalization of these invariants to elliptic and
hyperbolic equations were derived by Cotton (for the historical notes and
recent developments one may refer \cite{I:2,I:3}). Recently, renewal of interest
has been initiated to study the differential invariants of the equivalence
algebra of certain multidimensional linear PDEs as well as nonlinear PDEs
\cite{JM,I:4,Tr:1,ISA,ITV:1,ITV:2,Tr:2,TTV:1,TTV:2,IS,Dzh,I:6,SI,Tr:3,TT:1}.
\par
In this paper we explore both equivalence transformations and some of
their differential invariants for the eq.~(\ref{nlsorg}).  More
specifically by treating the functions $f,g$ and $h$ as arbitrary
parameters we explore the most general nonlinear PDE of the form
(\ref{nlsorg}) which can be transformed to the standard NLS.
Through this analysis we bring out a family of integrable variable
coefficient NLS equations that can be mapped to the standard NLS.
We also construct the transformation that connects the variable
coefficient NLS to the standard NLS.
\par
In the form of these equations   fall several cases  of cofficient
variable nonlinear Schr\"{o}dinger equations, in particular we
recall here that a wide subclass of equations considered in
\cite{yu} belongs to the class (\ref{nlsorg}), while this last one
could be considered a specialization of the socalled VCNLSE
considered by P. Winternitz and L.Gagnon in \cite{WG}.
\par
The plan of the paper is as follows.  In the following Sec.~2,  we study
the equivalence transformations of Eqs.~(\ref{nlsorg}).  In Sec.~3, we
investigate the invariants associated with (\ref{nlsorg}) and
show that the latter admits first order differential invariants.  As
an application of these differential invariants, in Sec.~4, we derive
the functional forms of $f$, $g$ and $h$ which characterize
the subclass of Eqs.~(\ref{nlsorg}) that  can be transformed
to a standard NLS Eq.~(\ref{snls}).  We present our
conclusions in Sec.~5.

\section{Equivalence transformations}

In order to study the equivalence transformations of Eqs.~(\ref{nlsorg}),
we rewrite the complex
equation (\ref{nlsorg}) as a system of real equations
by introducing a transformation $u=v+i\,w$, that is,
\numparts
%\label{nls1}
\begin{eqnarray}
v_t+f(t,x)w_{xx}+g(t,x)(v^2+w^2)w+h(t,x)w = 0, \label{nls2} \\
w_t-f(t,x)v_{xx}-g(t,x)(v^2+w^2)v-h(t,x)v = 0. \label{nls3}
\end{eqnarray}
\endnumparts
An equivalence transformation of the system
(\ref{nls2})-(\ref{nls3}) is  a non degenerate change of variables
from $(t,x,v,w)$ to $(\hat{t},\hat{x},\hat{v},\hat{w})$ and
transforming the equation of the form (\ref{nls2})-(\ref{nls3})
into another system of the same form but with different functions
$\hat{f}(\hat{t},\hat{x}),\; \hat{g}(\hat{t},\hat{x})$ and $
\hat{h}(\hat{t},\hat{x})$.  The equivalence transformations
for our system (\ref{nls2})-(\ref{nls3}) is obtained by making use of the Lie's
infinitesimal criterion\cite{OV}.  However, in the case of
the infinitesimal equivalence
generator we demand not only the invariance of
(\ref{nls2})-(\ref{nls3}) but also the invariance of the so called {\it
auxiliary conditions} in the {\it
augmented space} $(t,x,v,w,f,g,h)$.  In other words one needs to consider the
following conditions also, in addition to the ususal invariance conditions,
\begin{equation}
f_v = f_w = 0, \;\; g_v = g_w = 0, \;\; h_v = h_w = 0, \label{aux}
\end{equation}
which characterize the functional dependence of the functions
$f,\,g$ and $h$.

Let us consider the one-parameter group of equivalence
transformations, $G_{\cal E}$, in the {\it augmented space}
$(t,x,v,w,f,g,h)$ given by,
\numparts
%\label{trans}
\begin{eqnarray}
\hat t &=& t + \varepsilon \; \xi^1 (t,x,v,w) + {\cal O}(\varepsilon^2),
\label{trans1} \\
\hat x &=& x + \varepsilon \; \xi^2 (t,x,v,w) + {\cal O}(\varepsilon^2),
\label{trans2} \\
\hat v &=& v + \varepsilon \; \eta^1 (t,x,v,w) + {\cal O}(\varepsilon^2),
\label{trans3} \\
\hat w &=& w + \varepsilon \; \eta^2 (t,x,v,w) + {\cal O}(\varepsilon^2),
\label{trans4} \\
\hat f &=& f + \varepsilon \; \nu^1 (t,x,v,w,f,g,h) + {\cal O}(\varepsilon^2),
\label{trans5} \\
\hat g &=& g + \varepsilon \; \nu^2 (t,x,v,w,f,g,h) + {\cal O}(\varepsilon^2),
\label{trans6} \\
\hat h &=& h + \varepsilon \; \nu^3 (t,x,v,w,f,g,h) + {\cal O}(\varepsilon^2),
\label{trans7}
\end{eqnarray}
\endnumparts
where $\varepsilon$ is the infinitesimal group parameter.  The
vector field associated with the infinitesimal equivalence transformations
(\ref{trans1})-(\ref{trans7}) can be written as
\begin{equation}
Y=\xi^1\der{}{t}+\xi^2\der{}{x}+\eta^1\der{}{v}+\eta^2\der{}{w}
+\nu^1\der{}{f}+\nu^2\der{}{g}+\nu^3\der{}{h}. \label{EQ}
\end{equation}
Since (\ref{nls2})-(\ref{nls3}) involve second derivatives, we need
to consider second prolongation of the operator $Y$. Before
proceeding further we introduce the notation
\begin{eqnarray}
&& (f^1,f^2,f^3)\equiv (f,g,h), \quad
(x^1,x^2)\equiv(t,x),\quad(y^1,y^2)\equiv(v,w),  \nonumber \\
&& y^i_j=\der{y^i}{x^j},\quad
y^i_{jk}=\dder{y^i}{x^j}{x^k}, \quad j,i,k=1,2.
\label{notat}
\end{eqnarray}
With the above notations the second prolongation of the operator $Y$ can be
written as
\begin{eqnarray}
&&Y^{(2)}=Y+\zeta^i_j\der{}{y^i_j}+\zeta^i_{jj}\der{}{y^i_{jj}} +
\tilde\omega^r_{j}\der{}{f^r_{x^j}} +
\bar{\omega}^r_{j}\der{}{f^r_{y^j}}, \quad r=1,2,3,
\label{sp}
\end{eqnarray}
where the coefficients $\zeta_j^i$ and $\zeta^i_{jj}$ are given by
\begin{eqnarray}
\zeta^i_j=D_j\eta^i-y^i_kD_j\xi^k, \qquad
\zeta^i_{jj}=D_j\zeta^i_j-y^i_{jk}D_j\xi^k,  \nonumber
\end{eqnarray}
with
\begin{equation}
D_j=\der{}{x^j}+y^i_j\der{}{y^i}+y^i_{jk}\der{}{y^i_k}. \nonumber
\end{equation}
The remaining coefficients in (\ref{sp}) are obtained through the
following prolongation formula
\begin{equation}
\tilde \omega^r_{j}=\tilde D_{j} (\nu^r)-f^r_{x^k}\tilde D_{j} (\xi^k)
-f^r_{y^i}\tilde D_{j} (\eta^i),  \label{omega}
\end{equation}
where
\begin{equation}
\tilde D_{j}=\der{}{x^{j}}+f^r_{x^{j}}\der{}{f^r},\;\;
\bar D_{j}=\der{}{y^{j}}+f^r_{y^{j}}\der{}{f^r}.
\end{equation}
The invariance of Eqs.~(\ref{nls2})-(\ref{nls3}) under the
one-parameter group of equivalence transformations
(\ref{trans1})-(\ref{trans7}) can be written as \cite{OV}
\begin{eqnarray}
\fl \quad \quad
&&Y^{(2)} \left(v_t+f(t,x)w_{xx}+g(t,x)(v^2+w^2)w+h(t,x)w \right) = 0,
\label{invcond1} \\[1ex]
\fl \quad \quad &&Y^{(2)}
\left(w_t-f(t,x)v_{xx}-g(t,x)(v^2+w^2)v-h(t,x)v \right) = 0,
\label{invcond2} \\[1ex]
\fl \quad \quad &&Y^{(2)} \left(f_v\right) = Y^{(2)}
\left(f_w\right) = Y^{(2)} \left(g_v\right) = Y^{(2)}
\left(g_w\right) = Y^{(2)} \left(h_v\right) = Y^{(2)},
\left(h_w\right) = 0, \label{invcond3}
\end{eqnarray}
under the constraints that the variables $v$, $w$, $f$, $g$ and
$h$ have to satisfy the equations  (\ref{nls2})-(\ref{nls3}) and
(\ref{aux}).

 Substituting the second prolongation (\ref{sp}) into
(\ref{invcond1})-(\ref{invcond3}) and solving the resultant
equations ({\it determining system}) we find
%under constraints that $v$ and $w$ must satisfy Eqs. (\ref{nls1})
%and $f$,  $g$ and $h$ must satisfy (\ref{aux}) we
%get, following the well known algorithm, the {\it determining system},
%that allows to find infinitesimal components of $Y$
\begin{eqnarray}
\fl \quad \quad
\xi^1  = \varphi(t),\;\; \xi^2 = \psi(x),\;\;
\eta^1 = \left(a+\frac{\psi_x}{2}\right)v,\;\;
\eta^2  = \left(a+\frac{\psi_x}{2}\right)w,
\nonumber\\
\fl \quad \quad
\nu^1  = (2\psi_x -\varphi_t)f,\;\;
\nu^2  = -(\psi_x+\varphi_t+2\,a)g,\;\;
\nu^3  = -(h\varphi_t+\frac{1}{2}f\psi_{xxx}), \label{symm}
\end{eqnarray}
where $a$ is an arbitrary constant and $\varphi(t)$ and $\psi(x)$
are the arbitrary functions of $t$ and $x$ and subscripts denote
partial derivatives.

The associated equivalence algebra ${\cal E}$ is an infinite dimensional one
and is generated by the operators
\begin{eqnarray}
\fl \quad \quad
Y_{a} & = & v\frac{\partial}{\partial v}
                   + w\frac{\partial}{\partial w}
           -2g\frac{\partial}{\partial g},
\label{sym1} \\
\fl \quad \quad
Y_{\varphi} & = & \varphi\frac{\partial}{\partial t}
                  - f\varphi_t\frac{\partial}{\partial f}
                  - g\varphi_t\frac{\partial}{\partial g}
                  - h\varphi_t\frac{\partial}{\partial h},
\label{sym2}  \\
\fl \quad \quad
Y_{\psi} & = & \psi\frac{\partial}{\partial x}
                  +\frac{1}{2}v\psi_x\frac{\partial}{\partial v}
                  +\frac{1}{2}w\psi_x\frac{\partial}{\partial w}
                  +2f\psi_x\frac{\partial}{\partial f}
                  -g\psi_x\frac{\partial}{\partial g}
                  -\frac{1}{2}f\psi_{xxx}\frac{\partial}{\partial h}.
\label{sym3}
\end{eqnarray}

\section{Differential invariants of the equivalence algebra}

Differential invariants of order $n$ of the equivalence algebra
${\cal E}$ are not only
functions of the independent variables $t$ and $x$ but also
functions of $f$, $g$, $h$ and their derivatives upto to the maximal
order $n$ and  invariant with respect to the equivalence operator
$Y$.

\subsection{Differential invariants of order zero}

First let us seek differential invariants of order zero of the form
\begin{equation}
J = J(t,x,f,g,h).
\label{zein1}
\end{equation}
Applying the invariant test $Y(J) = 0$ to the operators $Y_a$,
$Y_{\varphi}$ with $\varphi=1$ and  $Y_{\psi}$ with $\psi=1$ we
find $J = J(f,h)$.  Since $\psi_x$ and $\psi_{xxx}$ are functionally
independent the invariant test $Y_{\psi}(J) =0$
provides us the following two equations
\begin{equation}
\frac{\partial J}{\partial f} = 0, \qquad
\frac{\partial J}{\partial h}= 0.
\label{zein3}
\end{equation}
As a result Eqs.~(\ref{nls2})-(\ref{nls3}) do not admit any
differential invariant of order zero.  In the following we seek
higher order differential invariants, if any.

\subsection{Differential invariants of first order}

To obtain differential invariants of first order
\begin{equation}
J = J(t,x,f,g,h,f_t,f_x,g_t,g_x,h_t,h_x),
\label{fiin1}
\end{equation}
in which the invariant includes both the spatial and time
derivatives of the functions $f,g$ and $h$, we consider the first
prolongation of the operator $Y$ given by
\begin{equation}
{Y^{(1)}} = Y + \tilde\omega^r_{j}\der{}{f^r_{x^j}}, \label{p}
\end{equation}
where the coefficients
$\tilde\omega^r_{j}$ can be constructed from Eq.~(\ref{omega}).
The explicit forms of the first prolongations of the generators $Y_a^{(1)}$,
$Y_{\varphi}^{(1)}$ and $Y_{\psi}^{(1)}$ are given by,
\begin{eqnarray}
\fl \quad \quad
 Y_a^{(1)} & = & v\frac{\partial}{\partial v}
           + w\frac{\partial}{\partial w}-2g\frac{\partial}{\partial g}
       -2g_x\frac{\partial}{\partial g_x}
       -2g_t\frac{\partial}{\partial g_t},
\label{p1} \\[1ex]
\fl \quad \quad
 Y_{\varphi}^{(1)} & = & \varphi\frac{\partial}{\partial t}
            -f\varphi_t\frac{\partial}{\partial f}
        -g\varphi_t\frac{\partial}{\partial g}
        -h\varphi_t\frac{\partial}{\partial h}
         -f_x\varphi_t\frac{\partial}{\partial f_x}
         -(2f_t\varphi_t+f\varphi_{tt})\frac{\partial}{\partial f_t},
\nonumber \\
\fl \quad \quad \quad
& &   -g_x\varphi_t\frac{\partial}{\partial g_x}
             -(2g_t\varphi_t+g\varphi_{tt})\frac{\partial}{\partial g_t}
     -h_x\varphi_t\frac{\partial}{\partial h_x}
             -(2h_t\varphi_t+h\varphi_{tt})\frac{\partial}{\partial h_t},
\label{p2}\\[1ex]
\fl \quad \quad
 Y_{\psi}^{(1)} & = & \psi\frac{\partial}{\partial x}
             +\frac{v}{2}\psi_x\frac{\partial}{\partial v}
             +\frac{w}{2}\psi_x\frac{\partial}{\partial w}
             +2f\psi_x\frac{\partial}{\partial f}
             -g\psi_x\frac{\partial}{\partial g}
              -\frac{f}{2}\psi_{xxx}\frac{\partial}{\partial h}
\nonumber\\
\fl \quad \quad \quad
& &         +(2f\psi_{xx}+f_x\psi_x)\frac{\partial}{\partial f_x}
            +2f_t\psi_x\frac{\partial}{\partial f_t}
            -(g\psi_{xx}+2g_x\psi_x)\frac{\partial}{\partial g_x}
\nonumber\\
\fl \quad \quad  \quad
& &         -g_t\psi_x\frac{\partial}{\partial g_t}
            -(\frac{f}{2}\psi_{xxxx}+\frac{f_x}{2}\psi_{xxx}+h_x\psi_x)
        \frac{\partial}{\partial h_x}
        -\frac{f_t}{2}\psi_{xxx}\frac{\partial}{\partial h_t}.
\label{p3}
\end{eqnarray}
The differential invariant test concerned with the function
$J$ given by (\ref{fiin1}) reads
\begin{equation}
Y_{\varphi}^{(1)}(J) = 0, \qquad
Y_{\psi}^{(1)}(J) = 0, \qquad
Y_a^{(1)}(J) = 0.
\label{fiin7}
\end{equation}
Since the arbitrary
functions $\varphi$ and $\psi$ and their derivatives  are to be
treated functionally independent, Eqs. (\ref{fiin7}) can be split
into the following conditions:
\begin{equation}
\frac{\partial J}{\partial t} = 0,\;\;\;
\frac{\partial J}{\partial x} = 0,\;\;\;
\frac{\partial J}{\partial h_x} = 0,
\label{det1}
\end{equation}
\begin{equation}
f\frac{\partial J}{\partial h}+f_t\frac{\partial J}{\partial h_t} = 0,
\label{det2}
\end{equation}
\begin{equation}
 2f\frac{\partial J}{\partial f_x}-g\frac{\partial J}{\partial g_x} = 0,
 \label{det3}
\end{equation}
\begin{equation}
f\frac{\partial J}{\partial f_t}+g\frac{\partial J}{\partial g_t}
+h\frac{\partial J}{\partial h_t} = 0,
\label{det4}
\end{equation}
\begin{equation}
g\frac{\partial J}{\partial g}+g_x\frac{\partial J}{\partial g_x}
+g_t\frac{\partial J}{\partial g_t}  = 0,
\label{det5}
\end{equation}
\begin{equation}
\fl \quad \quad \quad
4f\frac{\partial J}{\partial f}-2g\frac{\partial J}{\partial g}
+2f_x\frac{\partial J}{\partial f_x}+
4f_t\frac{\partial J}{\partial f_t}
-4g_x\frac{\partial J}{\partial g_x}-
2g_t\frac{\partial J}{\partial g_t} = 0,
\label{det6}
\end{equation}
\begin{equation}
\fl \quad \quad \quad
f\frac{\partial J}{\partial f}+g\frac{\partial J}{\partial g}
+h\frac{\partial J}{\partial h}+f_x\frac{\partial J}{\partial f_x}
+2f_t\frac{\partial J}{\partial f_t}+
g_x\frac{\partial J}{\partial g_x}
+2g_t\frac{\partial J}{\partial g_t}+
2h_t\frac{\partial J}{\partial h_t} = 0.
\label{det7}
\end{equation}
We note that Eq.~(\ref{det6}) can be simplified, with the use of (\ref{det5}), to
\begin{equation}
2f\frac{\partial J}{\partial f}+f_x\frac{\partial J}{\partial f_x}
+2f_t\frac{\partial J}{\partial f_t}-g_x\frac{\partial J}{\partial g_x} = 0.
\label{det8}
\end{equation}
As a result it is sufficient to solve Eqs.
(\ref{det1})-(\ref{det5}), (\ref{det7}) and (\ref{det8}) instead of
Eqs. (\ref{det1})-(\ref{det7}).
\par
Eq.~(\ref{det1}) simplify the
differential invariant (\ref{fiin1}) to the form
\begin{equation}
J = J(f,g,h,f_t,f_x,g_t,g_x,h_t). \label{fiin9}
\end{equation}
Solving the characteristic equation associated with (\ref{det2}),
\begin{equation}
\frac{dh}{f} = \frac{dh_t}{f_t}, \label{fiin9a}
\end{equation}
we get
\begin{equation}
\lambda_1=\frac{hf_t}{f}-h_t. \label{fiin10}
\end{equation}
Hence the function $J$ becomes $J = J(f,g,f_t,f_x,g_t,g_x,\lambda_1)$.

Similarly from (\ref{det3}) we obtain
\begin{equation}
\lambda_2=\frac{g}{f}f_x+2g_x \label{fiin11}
\end{equation}
and so the differential invariant reduces to
\begin{equation}
J = J(f,g,f_t,g_t,\lambda_1,\lambda_2).  \label{fiin9b}
\end{equation}
Substituting (\ref{fiin9b}) into  (\ref{det4}) and simplifying the
resultant equation we arrive at
\begin{equation}
f\frac{\partial J}{\partial f_t}+g\frac{\partial J}{\partial g_t} =0,
\label{fiin12a}
\end{equation}
from which we get
\begin{equation}
\lambda_3=\frac{g}{f}f_t-g_t. \label{fiin12}
\end{equation}
As a consequence (\ref{fiin9b}) can be further reduced to
$J=J(f,g,\lambda_1,\lambda_2,\lambda_3)$.  Substituting the latter into
(\ref{det8}) we obtain
\begin{equation}
2f\frac{\partial J}{\partial f}-\lambda_2\frac{\partial J}{\partial \lambda_2}
=0, \label{fiin13}
\end{equation}
which upon integration yields the invariant
\begin{equation}
\beta = f\lambda_2^2.
\label{fiin14}
\end{equation}
Therefore, the differential invariant $J$ assumes the form
\begin{equation}
J = J(g,\lambda_1,\lambda_3,\beta).
\label{fiin15}
\end{equation}
Now substituting (\ref{fiin15}) into (\ref{det5}) we get
\begin{equation}
g\frac{\partial J}{\partial g}
+\lambda_3\frac{\partial J}{\partial \lambda_3}
+2\beta\frac{\partial J}{\partial \beta} = 0.
\label{fiin16}
\end{equation}
The characteristic equations associated with the PDE
(\ref{fiin16}) can be written as
\begin{equation}
\frac{dg}{g} = \frac{d\lambda_3}{\lambda_3} =
\frac{d\beta}{2\beta}.
\label{fiin17}
\end{equation}
The invariant associated with the Eq.~(\ref{fiin17}) can be easily
found to be of the form
\begin{equation}
\alpha_1 = \frac{\lambda_3}{g}, \qquad \alpha_2 = \frac{\beta}{g^2}.
\label{fiin18}
\end{equation}
At this point, we have
\begin{equation}
J = J(\lambda_1,\alpha_1,\alpha_2)
\label{fiin19}
\end{equation}
and Eq.~(\ref{det7}) left unsolved.  Substituting (\ref{fiin19}) into
(\ref{det7}) and simplifying it we arrive at
\begin{equation}
2\lambda_1\frac{\partial J}{\partial \lambda_1}+\alpha_1
\frac{\partial J}{\partial \alpha_1}+\alpha_2
\frac{\partial J}{\partial \alpha_2} = 0.
\label{fiin20}
\end{equation}
By integrating the characteristic equations associated with the
PDE (\ref{fiin20}), we arrive at the following result.

\noindent
{\bf Theorem 1}.
The general form of the first order differential invariants of Eqs.
(\ref{nls2})-(\ref{nls3}) (or Eq. (\ref{nlsorg})), when $\lambda_1 \ne 0$, is
\begin{equation}
J = J(\gamma_1,\gamma_2),
\label{fiin21}
\end{equation}
where $\gamma_1$ and $\gamma_2$ are two independent invariants and their
explicit forms read
\begin{eqnarray}
\gamma_1= \frac{\alpha_1^2}{\lambda_1}=\frac{(gf_t-g_tf)^2}{fg^2(hf_t-h_tf)}, \qquad
\gamma_2= \frac{\alpha_2^2}{\lambda_1}=\frac{(gf_x+2g_xf)^4}{fg^4(hf_t-h_tf)}.
\end{eqnarray}
If $\lambda_1=0$, the corresponding form of Eqs.
(\ref{nls2})-(\ref{nls3}) (or Eq. (\ref{nlsorg})) should be
considered separately. It is now easy to show that the cases
$\alpha_1=0$ and $\alpha_2=0$ are also exceptional. Therefore, we can
state that the equations $\lambda_1=0$, $\alpha_1=0$ and
$\alpha_2=0$ are invariant with respect to the equivalence algebra
${\cal E}$.

\section{Application of differential invariants}

As stated earlier, our motivation is to map an equation of the
form (\ref{nlsorg}) to the standard NLS equation through
equivalence transformations.  To do this we make use of the
differential invariant (or the invariant equations) which we
derived in the previous section.  In particular, we consider the
following equation as target one
\begin{equation}
i\hat u_{\hat t}+k_1\hat u_{\hat x \hat x}+k_2|\hat u|^2\hat u = 0,
\quad k_1,k_2 \ne 0, \label{nls}
\end{equation}
where $k_1$ and $k_2$ are real constants.  One can easily check that the
Eq.~(\ref{nls}) can be transformed to (\ref{snls}) easily.

We observe that Eq.~(\ref{nls}) does not have any first order
differential invariant, but admits invariant equations
$\lambda_1=0$, $\alpha_1=0$ and $\alpha_2=0$.  This observation
leads us to formulate a necessary condition for an equation
belonging to the class  (\ref{nlsorg}) that can be mapped through
an equivalence transformation of $G_{\cal E}$ into (\ref{nls}).
The condition is that the functions $f,g$ and $h$ must satisfy
the following three equations
\begin{equation}
g\,f_t-f\,g_t=0, \qquad h\,f_t-f\,h_t=0, \qquad   g\,f_x+2f\,g_x=0. \label{iv}
\end{equation}
The most general forms of $f,g$ and $h$ satisfying (\ref{iv}) are
\begin{equation}
\fl \quad \quad
f = f_0\frac{n(t)}{l^2(x)}, \quad g = g_0\,n(t)l(x), \quad  h=n(t)m(x),
\quad f_0,\;g_0,\;n(t),\;l(x)\ne 0,
\label{iv1}
\end{equation}
where $l(x)$, $m(x)$ and $n(t)$ are real functions, while $f_0$ and
$g_0$ are real constants.  The above forms fix Eq.~(\ref{nlsorg}) of the form
\begin{equation}
iu_t + f_0\frac{n(t)}{l^2(x)}u_{xx} + g_0n(t)l(x)|u|^2u + n(t)m(x)u = 0.
\label{nlss}
\end{equation}
Eq.~(\ref{nlss}) can be further transformed into
\begin{equation}
iu_{\hat t} + f_0\frac{1}{l^2(x)}u_{xx} + g_0l(x)|u|^2u + m(x)u = 0,
\label{nlsse}
\end{equation}
by means of the equivalence transformation of $G_{\cal E}$
\begin{equation}
\hat t = \int^{t}\varphi(s)\,ds,\quad \hat x = x, \quad \hat u = u,
\label{tr1}
\end{equation}
with $\varphi(t) = n(t)$.

To transform (\ref{nlsse}) further into (\ref{nls}) we need to
calculate the second order differential
invariants of (\ref{nlsse}) or its associated real system
\begin{eqnarray}
v_{\hat t}+f_0\frac{1}{l^2(x)}w_{xx}+g_0l(x)(v^2+w^2)w+m(x)w = 0, \label{nlsse1} \\
w_{\hat t}-f_0\frac{1}{l^2(x)}v_{xx}-g_0l(x)(v^2+w^2)v-m(x)v = 0. \label{nlsse2}
\end{eqnarray}
We look for the functions of the form
\begin{equation}
J = J(\hat t, x, v, w, l, m, l_x, m_x, l_{xx}, m_{xx}),
\end{equation}
which are invariant with respect to the infinitesimal equivalence
generators of (\ref{nlsse1})-(\ref{nlsse2}), with the auxiliary conditions
$l_{\hat t}=m_{\hat t}=0$,
\begin{equation}
\Upsilon = {\bar \xi}^1\partial _{\bar t} + {\bar \xi}^ 2\partial _x
+{\bar \eta^1}\partial _v + {\bar \eta^2}\partial _w
+ \mu^1\partial _l + \mu^2\partial _m \label{gener}
\end{equation}
and its appropriate prolongations.

Introducing additional change of variables
\begin{equation}
f = \frac{f_0}{l^2}, \quad g = g_0l, \quad  h = m
\end{equation}
and following the procedure adopted in \cite{TT:2} for the change of
variables from the old coordinates $ \xi^1$, $ \xi^2$, $ \eta^1$,
$\eta^2$, $\nu^1$, $\nu^2$, $\nu^3$ of the generator $Y$ to the
new coordinates ${\bar \xi}^1$, ${\bar \xi}^2$, ${\bar \eta^1}$,
${\bar \eta^2}$, $\mu^1$, ${\mu^2}$ of $\Upsilon$, we obtain
\begin{eqnarray}
\bar \xi^1  = -\frac{4}{3}\,a\,\hat t + a_0, \qquad \bar \xi^2 = \xi^2 =
\psi(x), \nonumber \\[1ex]
\bar \eta^1 = \eta^1  = \left(a+\frac{\psi_x}{2}\right)v, \qquad
\bar \eta^2  = \eta^2  = \left(a+\frac{\psi_x}{2}\right)w, \nonumber \\[1ex]
%\fl
\mu^1  = -\left(\psi_x+\frac{2}{3}\,a\right)l, \qquad
\mu^2  = -\frac{f_0}{2\,l^2}\,\psi_{xxx}+\frac{4}{3}\,a\,m, \nonumber
\end{eqnarray}
with $a_0$ an arbitrary constant.

Considering the prolongation formula
\begin{equation}
\Upsilon^{(2)} = \Upsilon + \omega^1_x \partial _{l_x} + \omega^2_x
\partial _{m_x} + \omega^1_{xx} \partial _{l_{xx}} + \omega^2_{xx} \partial _{m_{xx}},
\end{equation}
where
\begin{eqnarray}
&& \omega^1_x = \der{\mu^1}{x} + l_x\der{\mu^1}{l}
+ m_x\der{\mu^1}{m} - l_x \der{\bar \xi^2 }{x}, \nonumber \\[1ex]
&& \omega^2_x = \der{\mu^2}{x} + l_x\der{\mu^2}{l}
+ m_x\der{\mu^2}{m} - m_x \der{\bar \xi^2 }{x}, \nonumber \\[1ex]
&& \omega^1_{xx} = \der{\omega^1_x}{x} + l_x\der{\omega^1_x}{l}
+ m_x\der{\omega^1_x}{m} + l_{xx}\der{\omega^1_x}{l_x}
+ m_{xx}\der{\omega^1_x}{m_x} - l_{xx} \der{\bar \xi^2 }{x}, \nonumber \\[1ex]
&& \omega^2_{xx} = \der{\omega^2_x}{x} + l_x\der{\omega^2_x}{l}
+ m_x\der{\omega^2_x}{m} + l_{xx}\der{\omega^2_x}{l_x}
+ m_{xx}\der{\omega^2_x}{m_x} - m_{xx} \der{\bar \xi^2 }{x} \nonumber
\end{eqnarray}
and repeating the same procedure followed in Sec.~4 we find that
the system (\ref{nlsse1})-(\ref{nlsse2}) (or Eq.(\ref{nlsse}))
does not possess any second order differential invariants but
admits the following invariant equation
\begin{equation}
m + \frac{1}{2}\,f_0\left(\frac{3}{2}\,\frac{l_x^2}{l^4} -
\frac{l_{xx}}{l^3}\right) = 0. \label{eqinv}
\end{equation}

By observing that (\ref{eqinv}) is also an invariant equation of
(\ref{nls}), we can conclude that the necessary condition for the
Eq.~(\ref{nlsorg}) to be mapped into (\ref{nls}) through the
equivalence transformation of $G_{\cal E}$ is that the functions
$f,g$ and $h$ must be related by the following relations
\begin{equation}
f=f_0\frac{n(t)}{l^2(x)}, \quad g=g_0n(t)l(x), \quad
h=\frac{1}{2}\,f_0\,n(t)\left(\frac{l_{xx}}{l^3}
- \frac{3}{2}\,\frac{l_x^2}{l^4}\right). \label{inv2}
\end{equation}
Now, we verify whether the conditions (\ref{inv2}) are also sufficient or not.

To do this, let us consider the equivalence transformation of $G_{\cal E}$,
that is,
\begin{equation}
\fl\qquad \hat t = \int^{t}\varphi(s)\,ds,\quad \hat x = \int^{x}\psi(s)\,ds,
\quad \hat u(\hat t,\hat x) = u(t,x)\,\sqrt{\psi(x)},
\quad \psi(x)>0 \label{tr2}
\end{equation}
and substituting it into the equation
\begin{equation}
iu_t + f_0\frac{n(t)}{l^2(x)}u_{xx} + g_0n(t)l(x)|u|^2u
+ \frac{1}{2}\,f_0\,n(t)\left(\frac{l_{xx}}{l^3}
- \frac{3}{2}\,\frac{l_x^2}{l^4}\right)u = 0, \label{nlse}
\end{equation}
 we get
\begin{eqnarray}
&& \varphi(t)\,i \hat u_{\hat t}
+ f_0\frac{n(t)\,\psi^2(x)}{l^2(x)}\hat u_{\hat x \hat x}
+ g_0\frac{n(t)\,l(x)}{\psi(x)}|\hat u|^2\hat u \nonumber \\[1ex]
&& \qquad \quad + \,\frac{1}{2}\,f_0\,n(t)\left(-\frac{\psi_{xx}}{\psi^3}
+ \frac{3}{2}\,\frac{\psi_x^2}{\psi^4} + \frac{l_{xx}}{l^3}
- \frac{3}{2}\,\frac{l_x^2}{l^4}\right)\hat u = 0. \label{nlse1}
\end{eqnarray}
Eq.~(\ref{nlse1}), with $f_0 = k_1$ and $g_0 = k_2$, reduces to
(\ref{nls}) when the transformation (\ref{tr2}) satisfies the following
conditions:
\begin{equation}
\varphi(t) = n(t), \qquad \psi(x) = l(x). \quad\Box
\end{equation}
As a result we demonstrated the following statement.

\noindent
{\bf Theorem 2}.
An equation belonging to (\ref{nlsorg}) can be transformed into the nonlinear
Schr\"{o}dinger equation (\ref{nls}) by an equivalence transformation of
$G_{\cal E}$
if and only if the functions $f$, $g$ and $h$ are given by (\ref{inv2}).

\vskip5mm
\noindent
{\bf Example}.
Let us consider the equation
\begin{equation}
iu_t + k_1\,t^{\frac{1}{2}}\,x^2 \,u_{xx} + k_2\,t^{\frac{1}{2}}\,x^{-1}\,|u|^2u
+ \frac{1}{4}\,k_1\,t^{\frac{1}{2}}\,\,u = 0. \label{es}
\end{equation}
Taking into account that $\varphi(t) = t^{\frac{1}{2}}$ and $\psi(x)
= x^{-1}$.  Using the change of variables (\ref{tr2}) we get
\begin{equation}
\hat t = \frac{2}{3}\,t \sqrt{t},\quad \hat x = \log{x}
\quad \hat u(\hat t,\hat x) = \frac{u(t,x)}{\sqrt{x}} \label{tres}
\end{equation}
and rewriting $(t,x,u)$ in terms of $(\hat{t},\hat{x},\hat{u})$ we obtain
\begin{equation}
t = \left(\frac{3}{2}\,\hat t\right)^{\frac{2}{3}}, \quad x = e^{\hat x},
\quad u(t,x) = e^{\frac{\hat x}{2}}\,\hat u(\hat t,\hat x). \label{tresinv}
\end{equation}
Using (\ref{tresinv}) one can
transform (\ref{es}) into (\ref{nls}). Eq.~(\ref{nls}),
with $k_1=k_2=1$ admits the following solution
\begin{eqnarray}
\hat u(\hat t, \hat x)= \sqrt{2}a\exp\bigg[i\frac{V_e}{2}\hat{x}
+i\bigg(a^2-\frac{V_{e}^2}{4}\bigg)\hat{t}\bigg]
sech\bigg[a(\hat x-V_e\hat t-x_0)\bigg].
\nonumber \\
\label{sol}
\end{eqnarray}
Rewriting (\ref{sol}) in terms of the old variables (vide Eq.~(\ref{tres})) one
gets
\begin{eqnarray}
\fl u(t,x) =
\sqrt{2}a\sqrt{x}\exp\bigg[i\frac{V_e}{2}\log{x}+i\bigg(a^2-\frac{V_{e}^2}{4}
\bigg)\frac{2t\sqrt{t}}{3}\bigg]
sech\bigg[a(\log{x}-\frac{2}{3}\,V_e\,t \sqrt{t}-x_0)\bigg]
\end{eqnarray}
which is a solution of (\ref{es}) with $k_1=k_2=1$ .

\section{Conclusions} In this paper we have constructed the equivalence
transformations for the family of variable coefficient nonlinear
Schr\"{o}dinger equations~(\ref{nlsorg}). We have shown that the
equivalence algebra is an infinite dimensional one. From the
infinitesimal equivalence generators, by using the invariant test,
we have shown that the family ~(\ref{nlsorg}) does not admit zero
order differential invariants but first order ones.  As an
application of the invariant relations we have characterized the
most general NLS family of equations (\ref{nlsorg}) that can be
mapped to the standard NLS equation by an equivalence
transformation. The explicit form of this latter one has also been
given explicitly.

\section*{Acknowledgements}
One of us (MT) wish to thank the members of the Centre for Nonlinear
Dynamics, Bharathidasan University, for their warm hospitality.
The work of MS forms part of a Department of Science and Technology, Government of India, sponsored research project.
The work of MT and AV is supported by MIUR-COFIN 2003/05 through the project {\em Nonlinear Mathematical Problems of
Wave Propagation and Stability in Models of Continuous Media} by the University of Catania through
{\it Progetti di Ricerca di Ateneo} and by INdAM through G.N.F.M..

\end{document}